\documentclass[runningheads]{cl2emult}

\usepackage{makeidx}  
\usepackage{graphicx} 
\usepackage{subeqnar} 
\usepackage{multicol} 
\usepackage{cropmark} 
\usepackage{eso}      
\makeindex            

\usepackage{amssymb}
\usepackage{natbib}
\bibpunct{(}{)}{;}{a}{}{,}

\newcommand{\decdegmm}[2]{#1\mbox{$^\circ\mskip-6.6mu.\,$}#2}

\begin{document}

\title*{Mining the Digital Hamburg/ESO \protect\newline
        Objective-Prism Survey
}
\toctitle{Mining the Digital Hamburg/ESO Objective Prism Survey}
%
%
\titlerunning{Mining the Hamburg/ESO survey}
%
\author{Norbert Christlieb\inst{1}
\and Lutz Wisotzki\inst{2}
\and Dieter Reimers\inst{1}
}
\authorrunning{N. Christlieb et al.}
%
%
\institute{Hamburger Sternwarte, Universit\"at Hamburg, Gojenbergsweg 112,
  \protect\newline D-21029 Hamburg, Germany, [nchristlieb,dreimers]@hs.uni-hamburg.de
\and Institut f\"ur Physik, Universit\"at Potsdam, Am Neuen Palais 10, 
  \protect\newline D-14469 Potsdam, Germany, lutz@astro.physik.uni-potsdam.de
}

\maketitle              

\begin{abstract}
  We report on the exploitation of the stellar content of the Hamburg/ESO
  objective prism survey\index{Hamburg/ESO survey} by quantitative selection
  methods.
\end{abstract}

\section{Introduction}

\begin{figure}[p]
  \begin{center}
    \includegraphics[ clip=, height=4cm, width=11cm,
       bbllx=203, bblly=451, bburx=501, bbury=569]{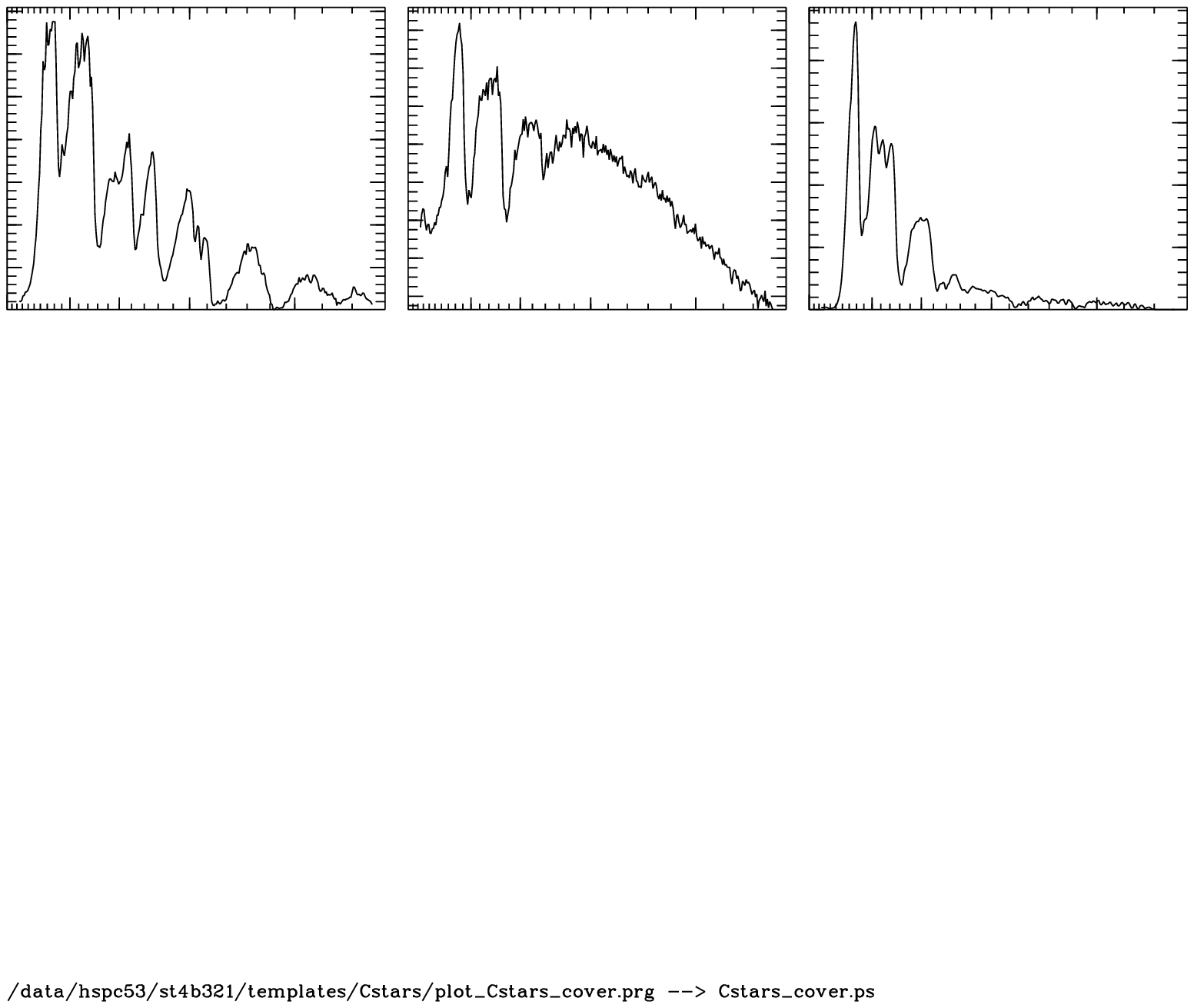}\\
    \includegraphics[ clip=, height=4cm, width=11cm,
       bbllx=203, bblly=451, bburx=501, bbury=569]{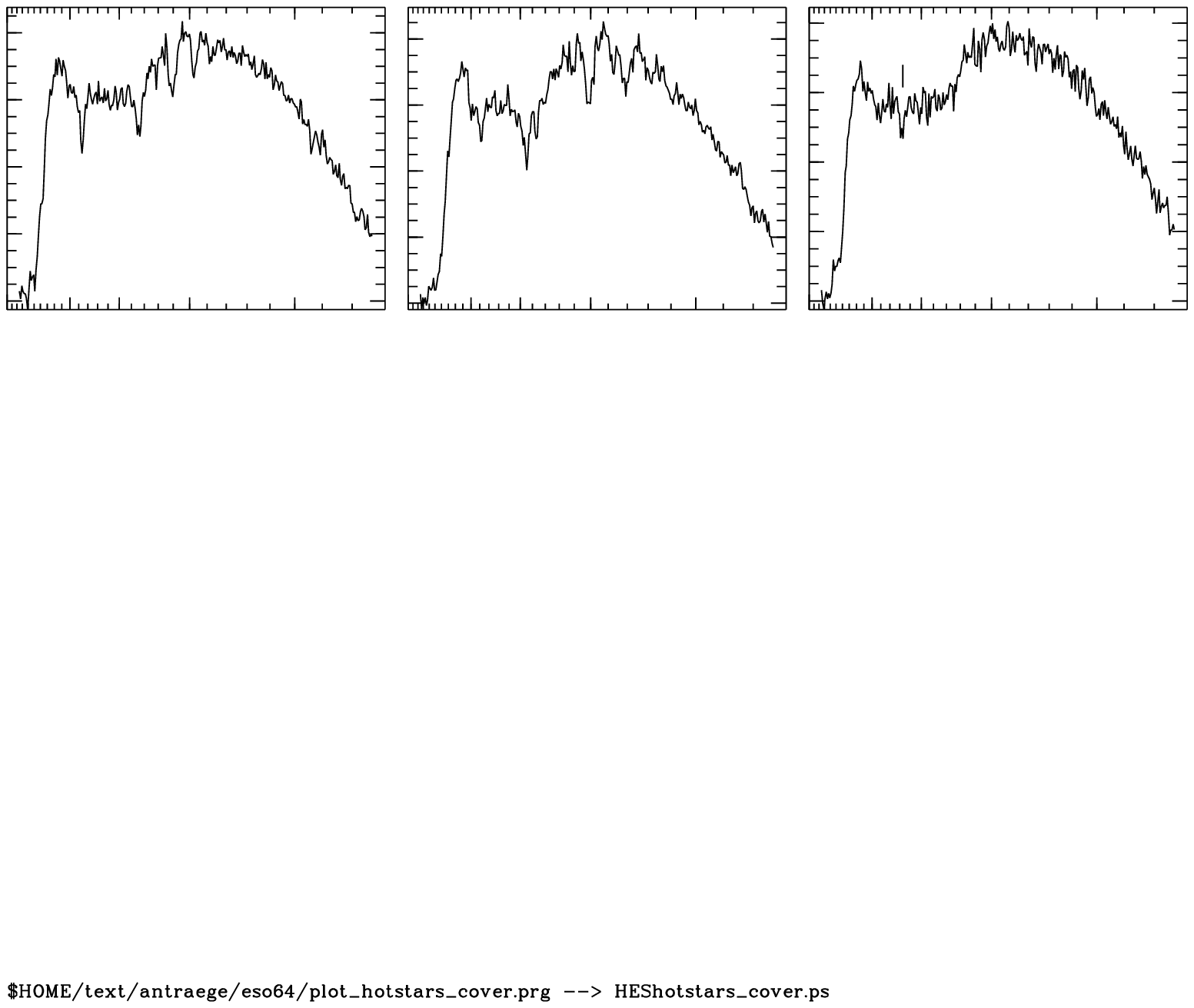}\\
    \includegraphics[ clip=, height=4cm, width=5.45cm,
       bbllx=52, bblly=451, bburx=200, bbury=569]{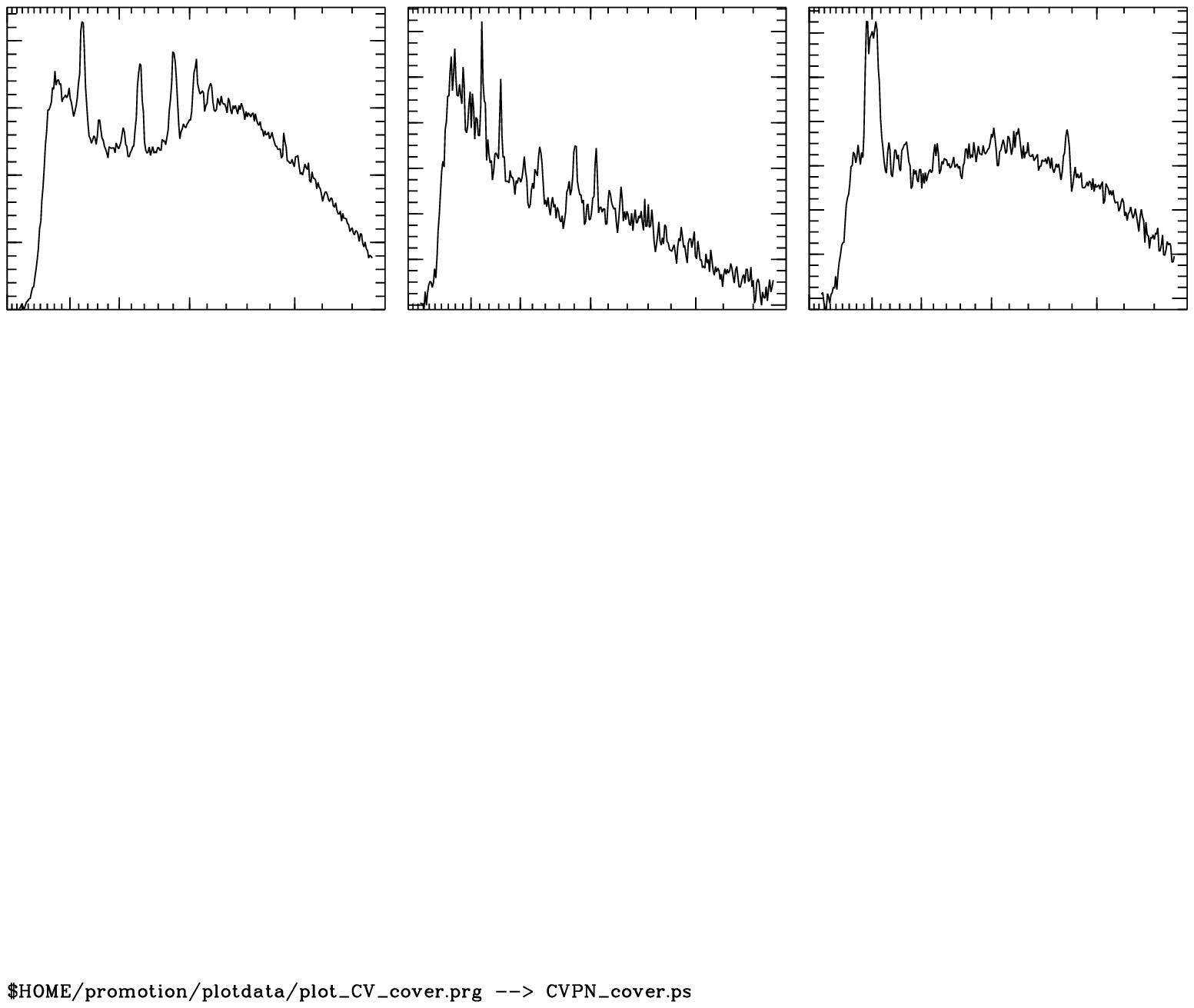}
    \includegraphics[ clip=, height=4cm, width=5.45cm,
       bbllx=354, bblly=451, bburx=501, bbury=569]{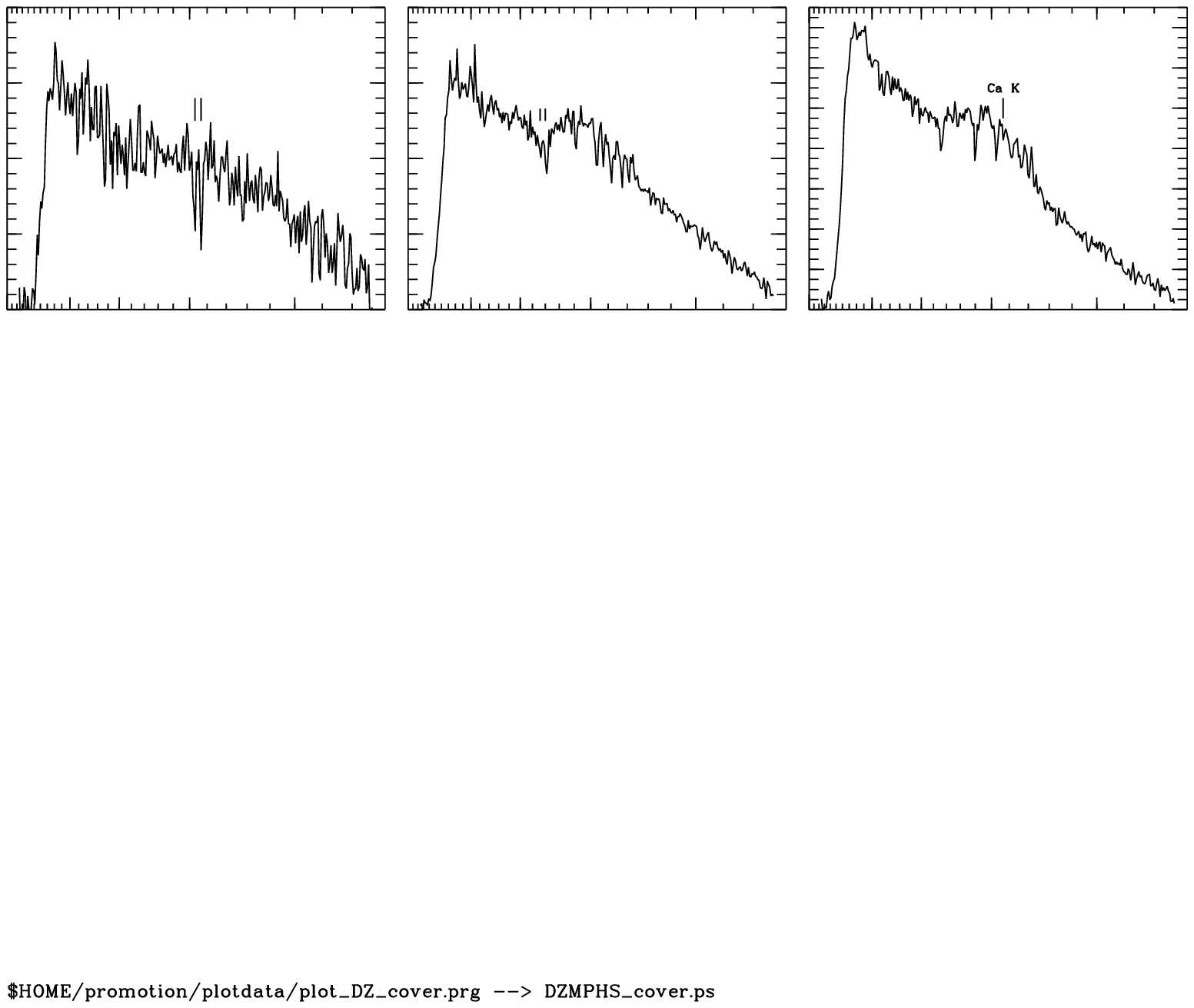}\\
    \includegraphics[ clip=, height=4cm, width=11cm,
       bbllx=52, bblly=442, bburx=351, bbury=569]{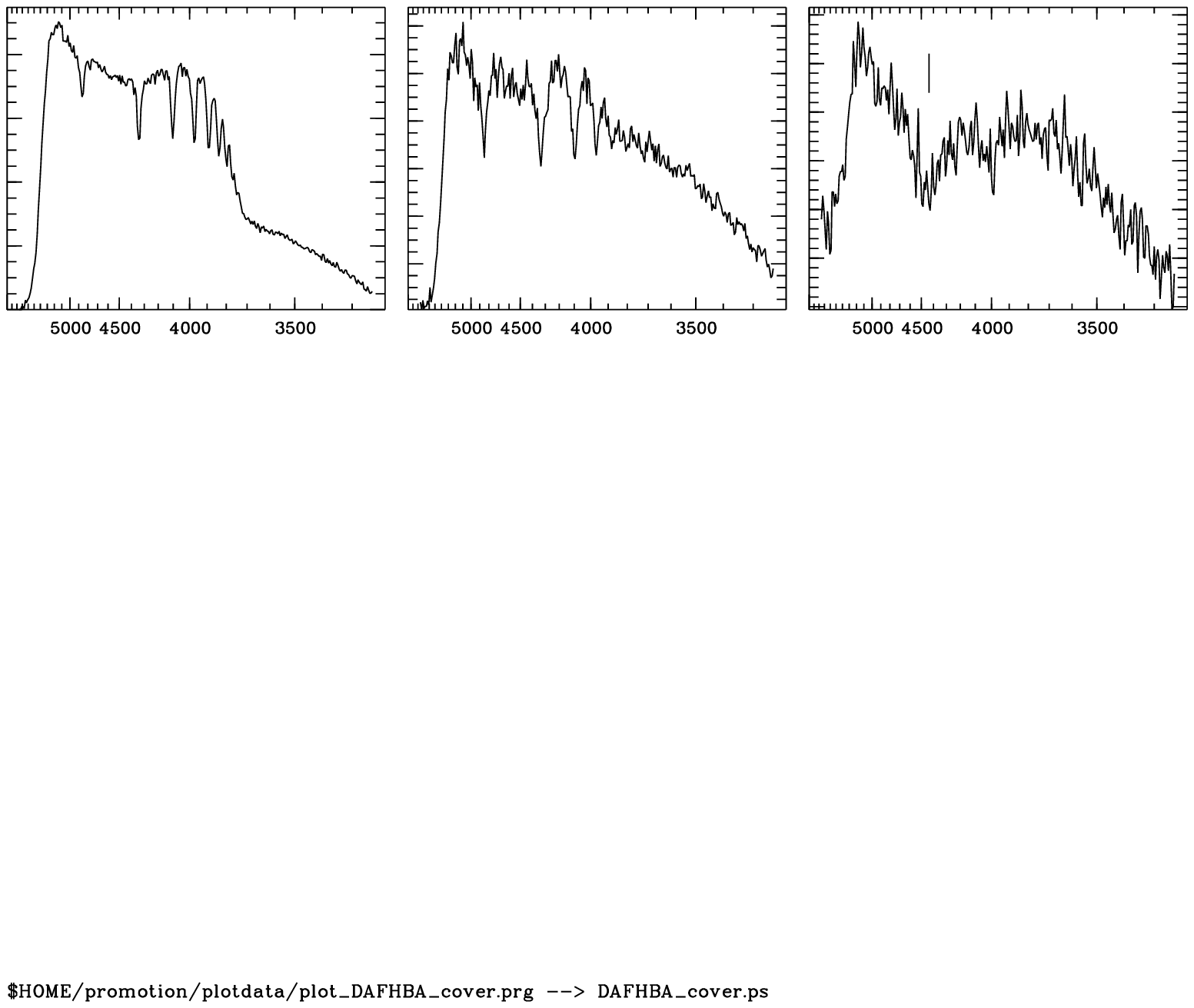}
    \caption{\label{star_examples} HES example spectra. Top panel:
       DQ white dwarf (left), cool carbon star (right); second panel: DB white
       dwarf, PG~1159 star; third panel: cataclysmic variable star, extremely
       metal-poor star, showing a very weak Ca~K line; lower panel: FHB/A
       star, cool DA white dwarf. Ordinates are photographic densities, abscissae
       wavelengths in {\AA}ngstr"om. Note that wavelength is \emph{de}creasing
       from left to right. The sharp cutoff at $\sim 5400$\,{\AA} is due to
       the IIIa-J emulsion sensitivity cutoff (``red edge'')}
  \end{center}
\end{figure}

The Hamburg/ESO survey \citep[HES; ][]{hespaperI,heshighlights,hespaperIII}
covers the total southern ($\delta < \decdegmm{2}{5}$) extragalactic
($|b|\gtrsim 30^{\circ}$) sky in the magnitude range $13.0 \gtrsim B_J \gtrsim
17.5$. It is primarily aiming at finding bright quasars. However, at its
spectral resolution of typically 15\,{\AA} FWHM at H$\gamma$, it is also
possible to efficiently select an abundance of interesting \emph{stellar}
objects. These include, e.g., metal-poor halo field stars, carbon stars,
cataclysmic variable stars (CVs), white dwarfs (WDs), subdwarf B stars (sdBs),
subdwarf O stars (sdOs), and field horizontal branch A- and B-type stars
(FHB/A).  Example spectra of some of these stars are displayed in Fig.
\ref{star_examples}.

\begin{figure}[htbp]
  \begin{center}
    \includegraphics[clip=, width=11.6cm, bbllx=48, bblly=420,
      bburx=527, bbury=740]{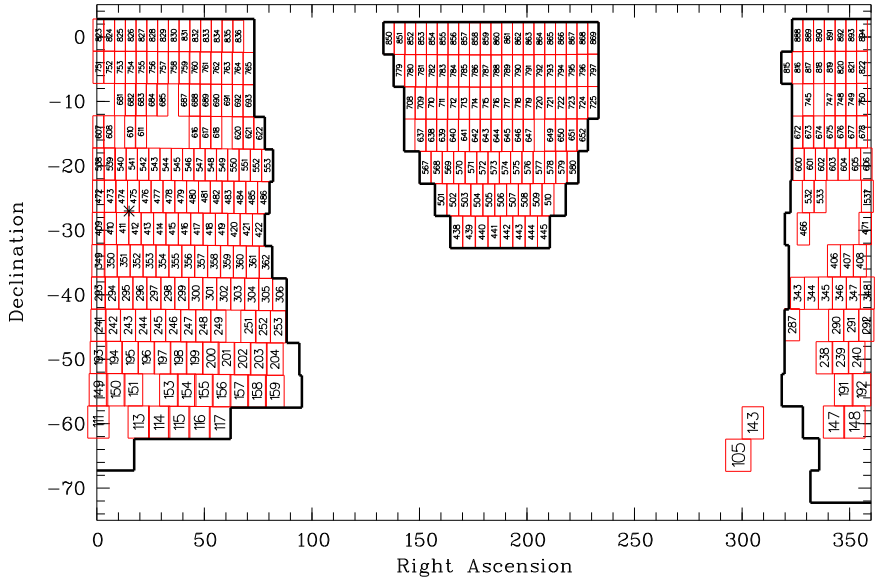}
    \caption{\label{stellar_hesarea} Definition of HES area (framed) and
      numbers of ESO/SERC fields in which the exploitation of the stellar
      content of the HES is currently carried out}
  \end{center}
\end{figure}

\cite{Christlieb:2000} has developed quantitative object selection methods,
such as automatic classification\index{Automatic spectral classification}, for
the systematic exploitation of the stellar content of the HES. In this paper
we describe the methods of automatic classification used, and report on
results obtained so far.

\section{Automatic spectral classification}\label{sect:autoclass}

Each of the $\sim 10$ million HES spectra can be represented by a feature
vector $\vec{x}$. A number of features are automatically detected in the
extracted and wavelength calibrated HES spectra
\citep{HESStarsI}.\index{Automatic feature detection} For stellar work, the
available features include equivalent widths of stellar absorption and
emission lines, line indices for $C_2$ and CN bands, principal components of
continua, broad band ($U-B$, $B-V$) and intermediate band (Str\"omgren $c_1$)
colours. The colours can be derived directly from HES spectra with accuracies
of $\sigma_{U-B}=0.092$\,mag, $\sigma_{B-V}=0.095$\,mag,
$\sigma_{c_1}=0.15$\,mag.

The goal of automatic classification in the HES is to identify objects of a
certain class in the large data base. That is, we want to construct a
\emph{decision rule} which allows to assign a spectrum with feature vector
$\vec{x}$ to one of the $n_c$ classes $\Omega_j$, $j=1\dots n_c$, defined in
the specific classification context. This is called \emph{supervised}
classification\index{Supervised classification}, as opposed to
\emph{un-supervised} classification, where the aim is to group objects into
classes not defined before the classification process.

For supervised classification a learning sample is always needed. For
our purposes, we define a learning sample to be a set of $n_{ls}$ objects for
which the feature vectors are known,
\begin{displaymath}
  \{\vec{x}\} = (\vec{x}_1,\dots,\vec{x}_{n_{ls}}),
\end{displaymath}
and for which the real classes are known. The real classes can be
defined e.g. by grouping a set of objects according to their stellar
parameters (e.g. $T_{\mbox{\scriptsize eff}}$, $\log g$, [Fe/H]), or by
assigning classes to a set of spectra by comparison with reference objects.
With the help of a learning sample, information on the class-conditional
probability densities $p(\vec{x}|\Omega_j)$
can be gained. $p(\vec{x}|\Omega_j)d\!\vec{x}$ is the
probability to observe a feature vector in the range $\vec{x}\dots
\vec{x}+d\!\vec{x}$, given the class $\Omega_j$. Experience has shown
that in most HES applications it is appropriate to model
$p(\vec{x}|\Omega_j)$ by multivariate normal distributions.

In many applications of automatic spectral classification in the HES, it is
not possible to generate a large enough learning sample from \emph{real}
spectra present on HES plates. This is because usually the target objects
are very rare. Therefore, we have developed methods to generate
\emph{artificial} learning samples by simulations, using either model spectra,
or slit spectra \citep{HESStarsI}.

\subsection{Bayes' rule classification}\index{Bayes' rule classification}

Classification with Bayes' rule minimizes the total number of
misclassifications, if the true distribution of class-conditional
probabilities $p(\vec{x}|\Omega_i)$ is used \citep{Hand:1981,Anderson:1984}.
Using Bayes' theorem,
\begin{displaymath}
  P(\Omega_i|\vec{x})=\frac{P(\Omega_i)p(\vec{x}|\Omega_i)}
  {\sum\limits_{\forall i}P(\Omega_i)p(\vec{x}|\Omega_i)},
\end{displaymath}
posterior probabilities $P(\Omega_i|\vec{x})$ can be calculated. A spectrum of
unknown class, with given feature vector $\vec{x}$, can then be classified
using Bayes' rule:
\begin{description}
\item[Bayes' rule:] \emph{Assign a spectrum with feature vector $\vec{x}$ to
    the class with the highest posterior probability $p(\Omega_i|\vec{x})$.}
\end{description}

The achievable accuracy of any automatic spectral classification algorithm
always depends on the signal-to-noise ratio ($S/N$) of the data used. In the
HES, the accuracies for spectra in the colour range $0.3<B-V<0.7$, with
$S/N>10$ (typically corresponding to $B<16.5$), are $\sigma_{\mbox{\scriptsize
    Teff}}<400$\,K (or $<1.6$ MK types), $\sigma_{\log g}<0.68$\,dex (or
$<0.55$ luminosity classes) and $\sigma_{\mbox{\scriptsize
    [Fe/H]}}<0.68$\,dex.  The classification accuracy in [Fe/H] strongly
depends on [Fe/H] itself, and is much better than $0.68$\,dex for
$\mbox{[Fe/H]}>-2.0$.  For cooler stars ($T_{\mbox{\scriptsize eff}}<5200$\,K)
not yet covered by our learning sample, the accuracy of the luminosity
classification is expected to be lower, since the sensitivity of $c_1$ to
gravity is higher in hotter stars.
  

\subsection{Minimum cost rule classification}\index{Minimum cost rule classification}

In many of the classification problems arising in the HES it is desired to
compile a sample of objects of a specific class, or a specific set of
classes. In these cases, Bayes' rule is not appropriate, because we do not
want to minimize the total number of misclassifications, but the
misclassifications between the desired class(es) of objects, and the remaining
classes. Suppose we have three classes, A-, F-, and G-type stars, and we want
to compile a complete sample of A-type stars. Then only misclassifications
between A-type stars and F- and G-type stars (and vice versa) are of interest.
More specifically, misclassifications of A-type stars to F- and G-type stars
(leading to incompleteness) are least desirable when a complete sample shall
be compiled, and erroneous classification of F- and G-type stars as A-type
stars (resulting in sample contamination) can be accepted at a moderate rate.
Misclassifications between F- and G-type stars can be totally ignored, because
the target object type is not involved.

Classification aims like this can be realized by using a minimum cost rule.
Cost factors $r_{hk}$, with
\begin{equation}\label{Def_Verlustfaktoren}
  0 \le r_{hk} \le 1; \qquad h = 1,\dots,n_c;\quad k = 1,\dots,n_c,
\end{equation}
allow to assign \emph{relative weights} to individual types of
misclassifications.  The cost factor $r_{hk}$ is the relative weight of a
misclassification from class $\Omega_h$ to class $\Omega_k$.

Suppose we have an object of unknown class, with feature vector $\vec{x}$. We
ask how large the cost is, if it belongs to class $\Omega_h$, and would be
assigned to class $\Omega_k$, $h\not= k$. The cost $C_{h\to k}(\vec{x})$ is:
\begin{equation}
C_{h\to k}(\vec{x}) = r_{hk}P(\Omega_h|\vec{x}).\label{Vhk}
\end{equation}
We do not know to which of the possible classes $\Omega_h$, $h = 1,\dots,n_c$,
the object actually belongs. Therefore, we estimate the expected cost
$C_k(\vec{x})$ for assigning an object with feature vector $\vec{x}$ to the
class $\Omega_k$ by computing the following sum of costs:
\begin{eqnarray}
  C_k(\vec{x}) &=& \sum\limits_{h=1\atop h\not=k}^{n_c}
                   C_{h\to k}(\vec{x})\nonumber\\
               &=& \sum\limits_{h=1\atop h\not=k}^{n_c}
               r_{hk}P(\Omega_h|\vec{x}).\label{V_k}
\end{eqnarray}

\noindent Now we can formulate the minimum cost rule, which minimizes the
total cost \citep{Hand:1981}.
\begin{description}
\item[Minimum Cost Rule:] {\em Assign an object with feature vector $\vec{x}$
    to the class $\Omega_k$ with the lowest expected cost $C_k(\vec{x})$.}
\end{description}

If the cost factors are chosen such that $r_{hk}\equiv\delta_{hk}$, the
minimum cost rule classification is identical to classification according to
Bayes' rule. In this case the cost for assigning the class $\Omega_k$ to a
spectrum with feature vector $\vec{x}$ is the probability that the object
belongs to one of the other classes $h\not= k$. This follows immediately from
(\ref{V_k}). If $r_{hk}\not=\delta_{hk}$, the total number of
misclassifications is \emph{not} minimized, so that the quality of a minimum
cost rule classification has to be evaluated by other criteria, such as sample
completeness for a given contamination.

\section{First results}\label{sect:results}

We briefly report on first results from the HES stellar work. More details can
be found in the paper series ``The stellar content of the Hamburg/ESO
objective prism survey'', which is currently being published in A\&A, and in
\cite{Christlieb:2000}.

\subsection{Metal-poor stars}\index{Metal-poor stars}

Spectroscopic follow-up observations of 58 candidate metal-poor halo stars
selected by automatic spectral classification in the HES showed that this
selection has a more than three times higher efficiency then the selection in
the only other spectroscopic wide angle survey for such stars, the so-called
HK survey \citep{BPSII,TimTSS}. The effective yield of turnoff stars with
$[$Fe/H$]<-2.0$ is 80\,\% in the HES, but only 22\,\% in the HK survey on
average. This is very remarkable considering the fact that the spectral
resolution of the HES ($\sim 10$\,{\AA} at Ca~K) is two times \emph{lower}
than in the HK survey ($\sim 5$\,{\AA}). The advantages of the HES are: (a)
broader wavelength coverage, (b) better quality of the spectra, and (c) 
automated candidate selection procedures, as opposed to visual inspection of
objective prism plates in the HK survey.

In spectroscopic follow-up campaigns of metal-poor stars carried out so far,
90 metal-poor stars were discovered; 11 are unevolved stars with
$\mbox{[Fe/H]}\le-3.0$.  Since in the HK survey 37 stars with
$\mbox{[Fe/H]}\le-3.0$ and $0.3<(B-V)_0<0.5$ were found, the sample of
unevolved, extremely metal-poor stars was already increased noticeably.  First
abundance analysis using high resolution spectra obtained with UVES, the
high-resolution spectrograph attached to VLT-UT2, was recently published
\citep{Depagneetal:2000}. We plan to use the multi-fiber spectrograph 6dF at
the UK Schmidt telescope to follow-up the thousands of metal-poor candidates
that were selected in the HES, in order to provide more of these interesting
targets for high resolution studies using 10\,m class telescopes.

\subsection{Carbon stars}\index{Carbon stars}

On the 329 HES plates used so far for stellar work (effective area $\sim
6\,400$\,deg$^2$, or 87\,\% of the HES), 351 carbon stars where identified.
The mean surface density detected by the HES hence is 0.055\,deg$^{-2}$, which
is almost a factor three higher than the surface density found by
\cite{Greenetal:1994} in their photometric CCD survey.  Moreover, the survey
of Green et al. is $\sim 1.5$\,mag \emph{deeper} than the HES
($V_{\mbox{\scriptsize lim}}\sim 16.5$ in the HES;
$V_{\mbox{\scriptsize lim}} \sim 18.0$ for the Green et al. survey).
This indicates that photometric carbon star surveys are highly incomplete.

We have obtained recent epoch CCD frames of most of the HES carbon stars.
Comparison with archival plate material available online is currently being
done to derive proper motions, and identify halo dwarf carbon stars in our
sample.

\subsection{White dwarfs}\index{White dwarfs}

One exciting application for the hundreds of new white dwarfs that were found
in the HES is testing the double-degenerate (DD) scenario for type Ia
supernoave (SN~Ia) progenitors, in which a binary system, consisting of two
white dwarfs of large enough mass, merges and produces a thermonuclear
explosion. Although a couple of DDs\index{Double degenerates} were identified
by radial velocity (RV) variations, past efforts have failed so far to
identify any SN~Ia progenitor systems among the DDs, which is being attributed
to too small sample sizes \citep{Maxted/Marsh:1999}. In a \emph{Large
  Programme} approved by ESO (P.I.: Napiwotzki), we use VLT-UT2/UVES to
observe WDs at two randomly chosen epochs, to find more DDs. We aim at
observing a total set of $\sim 1500$ DAs and DBs selected in the HES data
base, and taken from the literature. With a set of 224 spectra of 107 WDs
processed so far, 15 objects with RV variations were found (Napiwotzki 2000,
priv. comm.).  Follow-up observations to determine the orbital periods for
these systems will be carried out in the near future.


\clearpage
\addcontentsline{toc}{section}{Index}
\flushbottom
\printindex

\end{document}